\documentclass[10pt,superscriptaddress,aps,pra]{revtex4-2}
\usepackage[bookmarks=false,colorlinks=true]{hyperref}
\usepackage{amsmath,amsthm,amssymb}

\counterwithin*{equation}{section}

\begin{document}

\title{Parafermionic representation of Potts-based cluster chain}
\author{Tigran Hakobyan}
\email{tigran.hakobyan@ysu.am}
\email{hakob@yerphi.am}
\affiliation{Yerevan State University, 1 Alex Manoogian Street, Yerevan, 0025, Armenia}
\affiliation{Alikhanyan Natianal Laboratory, 2 Aliknanyan br. Street, Yerevan, 0036, Armenia}
\author{Raffi Varosyan}
\email{raffi.varosyan@edu.ysu.am}
\email{raffivarosyan88@gmail.com}
\affiliation{Yerevan State University, 1 Alex Manoogian Street, Yerevan, 0025, Armenia}

\begin{abstract}
The cluster chain with $\mathbb{Z}_p \times \mathbb{Z}_p$ symmetry-protected topological (SPT) order is decomposed
into two distinct bilinear parafermionic chains, each possessing intrinsic topological order.
These chains are formed by standard parafermions and time-reversal parafermions,
respectively. Each subsystem retains its own $\mathbb{Z}_p$ symmetry component,
which characterizes the total parity of constituent particles.
Their topological orders are inherited from the two SPT orders of the cluster model.
The transformations of particles under reflection, translation, and time reversal are derived. In the open chain,
four zero-energy parafermionic edge modes are identified, and their structure is analyzed.
For the closed system, the boundaries are twisted by the total parafermion parity.
It is shown that the open chain is reflection-invariant when the number of spins is even,
and $\mathcal{PT}$-invariant when the number of spins is odd.
Meanwhile, the closed model exhibits a symmetry characterized by an
antiunitary analog of the dihedral group.
\end{abstract}

\maketitle

\section{Introduction}
In the last decade, much attention has been paid to
the structure and properties of the quantum states of matter at
zero temperature. The interest is explained by rapid developments
in quantum technologies: quantum computers, quantum teleportation, etc.
\cite{book-wen}.
Quantum phases have a much richer structure compared to classical ones.
According to Landau theory, the classical phases are characterized
by the symmetry of the system, and a change or violation of symmetry
leads to a phase transition. However, due to the complexity of the wave function,
symmetry alone is not enough to characterize different quantum states of matter.
The hidden topological properties of the ground state enter into the game.
%
They are characterized  by nonlocal order parameters and
might also be uncovered by the appearance of the gapless excitations and anomalous symmetries at the edge.

Let us take a closer look at quantum matter, which has a unique, gapped ground state (in the bulk)
with short-range  entanglement. If, in addition, a certain symmetry exists, a single Landau
phase splits into several symmetry-protected topological (SPT) phases \cite{SPTwen,SPTpoll}. Two states
belong to the same SPT phase if they can be smoothly deformed into each other without closing
the energy gap and  without breaking the protecting symmetry.
Separable states formed by the tensor product of single-particle states
belong to the trivial phase.
The most familiar example that exhibits SPT order is the
topological insulator (in two dimension: the quantum spin Hall state) based on free
fermions \cite{kane,balents,kane3d,insul-rev}. Topological insulators are
 experimentally realized \cite{insul-exp}.
The SPT phase there is protected by time reversal
and  charge conservation symmetries.
Other examples are the antiferromagnetic spin-1 Heisenberg
chain studied by Haldane \cite{haldane83}  and its  slightly modified
version with exact ground state \cite{AKLT}.
The Haldane phase is protected by $SO(3)$ spin rotation symmetry \cite{SPTwen}.
It is protected also by any one of the following discrete symmetries:
the dihedral group of $\pi$-rotations about coordinate axes, time reversal, and chain reflection \cite{SPTpoll}.
%
%
In contrast to SPT states, the quantum
states of matter with intrinsic topological orders correspond to patterns of
long-range entanglement, which are robust against any local
perturbation that can break any symmetry \cite{topo,zoo-topo}. Topological phases have been
observed in
conventional superconductors \cite{super},  chiral spin  states \cite{topo}, and fractional quantum Hall states \cite{laughlin}.

The  cluster chain with $\mathbb{Z}_2\times\mathbb{Z}_2 $ symmetry \cite{son} is the simplest model
with nontrivial SPT order. Its trivial version corresponds to the standard Ising model.
The cluster  model is  used  in measurement-based quantum computation  \cite{rauss,measur-based}.
Recently, it has been extended to higher spins. The related system is characterized by
the  $\mathbb{Z}_p\times\mathbb{Z}_p$ SPT order \cite{santos,cluster,dipolar}. The trivial phase is described by the
 $p$-state Potts model.
Unlike the Ising case, the SPT Hamiltonian is inhomogeneous along the chain.
It splits into two homogeneous parts consisting of odd and even local terms,
which become similar to each other when the time reversal transformation is applied.

The cluster model is closely related to Kitaev's superconducting chain, which is one of the simplest fermion systems exhibiting nontrivial topological order. This system can be represented in terms of spinless Majorana fermions. Topological order manifests itself through zero-energy edge modes \cite{kitaev}. Recently, these modes have been observed experimentally \cite{poor-man}.
The multiple Majorana modes exhibit non-abelian braiding statistics, known as parastatistics \cite{ivanov}. This property, combined with a strong energy gap and the robustness of zero modes against local perturbations, makes them a promising tool for fault-tolerant topological quantum computing.
  %
%
The Jordan-Wigner mapping between Kitaev chains and cluster models reveals a deep connection between
their topological properties \cite{verresen}. Specifically, decoupled Kitaev chains are mapped to cluster-type models with expanded local stabilizers. Notably, the standard cluster chain corresponds to two decoupled Kitaev chains. Together, these chains are equivalent to the Su-Schrieffer-Heeger model \cite{SSH}.

In this article, a parafermion representation of a higher-spin cluster chain is derived, and
the topological  and symmetry properties of the resulting system are examined.
Parafermions are specific anyons that obey fractional statistics and can also be regarded
as an extension of spinless Majorana fermions. Recently, they have gained attention as a potential source
for qudits in high-dimensional quantum computers \cite{z4par}, which have become
increasingly popular \cite{qudits}.
These quasiparticles may play an even more significant role in quantum computation due to their braiding,
which enables the performance of entangling quantum gates \cite{loss,photonic}.
Compared to Majoranas, they adhere to fractional and position-dependent statistics.
Lattice parafermions emerge as quasiparticle excitations in clock models, much like
Majorana fermions arise in the Ising model \cite{FK,alcaraz}.
Their protected zero modes appear at the boundaries of specific one-dimensional
$\mathbb{Z}_p$-invariant spin systems \cite{fendley12,review}.
Recent experiments have demonstrated the feasibility of realizing parafermionic modes at the edges
of fractional quantum Hall systems at certain filling factors that align with parastatistics \cite{para-hall}.


We begin with the $N$-site SPT cluster chain with protecting $\mathbb{Z}_p \times \mathbb{Z}_p$ symmetry.
It is divided into two distinct parafermion chains,
each exhibiting an intrinsic topological order.
Compared to the $p=2$ case \cite{verresen}, the situation here is more intricate and entangled due to
the inhomogeneity between the odd and even local terms in the spin Hamiltonian,
as well as the inherent complexity of the parafermions.
A direct application of the spin-to-parafermion mapping results in a six-particle interaction
term in the Hamiltonian, which disrupts its symmetric structure.
We restore the original symmetric form with the inclusion of
time-reversal ($\mathcal{T}$) particle operators.
Consequently, the Hamiltonian becomes bilinear in particle operators,
except in the translationally invariant case, where an additional factor representing
a composite parafermion emerges at the boundary.
This factor is responsible for the diagonal $\mathbb{Z}_p$ symmetry and describes
 the total parity of all $2N$ particles.
It serves as a parafermionic analog to Dirac's chiral
gamma matrix
and manifests when the particles undergo transformations induced by lattice reflection
($\mathcal{P}$) and translation ($T$).

We establish that the parafermionic cluster model, along with the protecting symmetry,
zero-energy edge modes, and string order operators,
is divided into two distinct parts, each characterized by its own particles.
The first (odd) part is represented by the standard parafermions with index values
$i=0,3$ modulo 4, and their total parity corresponds to the $\mathbb{Z}^\text{even}_p$
 symmetry.
Meanwhile, the second (even) part is defined by the time-reversed parafermions
with $i=2,1$ modulo 4, and their total parity corresponds
to the $\mathbb{Z}^\text{odd}_p$ symmetry.
For the open chain, four independent edge parafermions are identified, which
generate conserved zero-energy modes.
Their structure depends on whether the chain length is odd or even.
This distinction arises from differences in symmetry: for odd
$N$, the Hamiltonian exhibits $\mathcal{PT}$ invariance, while for even
$N$, it displays $\mathcal{P}$ invariance.
In all cases, we associate two edge parafermions with the first chain and the
remaining two with the second chain.
The left and right invariants, which are responsible for symmetry fractionalization,
are expressed in terms of the edge parafermions.

We also consider periodic boundary conditions. In this case, along with
$\mathcal{P}$ invariance (when $N$ is even), an additional
$T\mathcal{T}$ symmetry (time-reversal combined with lattice translation) emerges.
This symmetry interchanges two subsystems.
As a result, the corresponding Hamiltonian becomes twisted by the composite parafermion,
whose eigenvalues are $p$-th roots of unity.
When restricted to the invariant sector associated with a specific eigenvalue,
the Hamiltonian acquires a bilinear form with twisted boundary conditions.

The article is organized as follows: In Sect.~\ref{sec:cluster},
the $p$-state cluster chain with open and periodic boundary conditions is briefly reviewed.
Specifically, the spin Hamiltonian, its nontrivial SPT phase with the $\mathbb{Z}_p \times \mathbb{Z}_p$
symmetry, the string order parameter, zero-energy boundary modes,
and symmetry fractionalization (for open chains) are described.
In Sect.~\ref{sec:para}, the direct and inverse Fradkin-Kadanoff transformations are
used to construct the spin representation for the symmetry generators and parafermion bilinears.
Subsequently, the parafermionic representation of the Potts-based cluster model is
derived for open boundary conditions. However, the related Hamiltonian includes unwanted sixth-order
terms in the particle operators.
Sect.~\ref{sec:native} focuses on the construction and analysis of the bilinear representation
of the same model and its invariant edge modes.
Firstly, the behavior of parafermions under the $\mathcal{P}$, $\mathcal{T}$, and $\mathcal{PT}$
transformations is examined in detail. The combined set of original and time-reversal particle
operators obey extended exchange relations that closely resemble the parafermion algebra.
Next, the even (unwanted) part of the cluster model is reformulated in terms of time-reversal parafermions,
resulting in the entire Hamiltonian acquiring a bilinear form.
Finally, the symmetry and structure of the invariant parafermions emerging at the chain edges
and the factorization of the system into two separate subchains are examined comprehensively.
In Sect.~\ref{sec:Tinv}, the transformation of the particle operators and their bilinears under
translation is discussed. The parafermionic representations of the closed $\mathbb{Z}_p$ cluster chain,
along with the related string order parameter, are constructed,
and the space-time symmetries are analyzed thoroughly.
The results are briefly summarized in the concluding section.

\section{Cluster chain based on Potts model}
\label{sec:cluster}

\paragraph*{Potts-based cluster chain with periodic boundary conditions.}
The cluster model is the simplest example that exhibits a nontrivial SPT order \cite{son,book-wen}.
Its ground state is used in measurement-based quantum
computation \cite{rauss}.
The system is unitary equivalent to the noninteracting Ising model. In one dimension, it
protects the $\mathbb{Z}_2\times \mathbb{Z}_2$ symmetry, which is
generated by the spin reflections applied independently on the even and odd sites.
In this section, we review an extended cluster chain which corresponds to the SPT version of the $p$-state Potts model
\cite{cluster,santos,dipolar}.

The single-site states of the considered model
can be parameterized by the elements of the cyclic group $\mathbb{Z}_p$.
For $p>2$, the odd and even local terms in the  Hamiltonian are given by different  three-spin couplings.
It is suitable to split the model into two components:
\begin{align}
\label{H}
&H = H_\text{odd} + H_\text{even},
\end{align}
which describe, respectively,  the odd and even subsystems:
\begin{align}
\label{Hodd}
&H_\text{odd} = -\frac{1}{2}\sum_{r=1}^{L} \left( Z_{2r}^+X_{2r+1}Z_{2r+2} + Z_{2r}X_{2r+1}^+Z_{2r+2}^+\right),
\\
\label{Heven}
&H_\text{even} = -\frac{1}{2}\sum_{r=1}^{L} \left(Z_{2r-1}X_{2r}Z_{2r+1}^+ + Z_{2r-1}^+X_{2r}^+Z_{2r+1}\right).
\end{align}
First, we consider the \emph{periodic} boundary conditions with an even number of  spins: $N=2L$.
Then, the modulo $N$ summation is supposed inside all indices.
Here,  $p$-state extensions ($p=2,3,\dots$) of the  standard Pauli matrices ($X=\sigma_x$, $Z=\sigma_z$) are introduced. They
obey the following algebra:
\begin{equation}
\label{ZX}
ZX=\omega XZ,
\qquad
X^p=Z^p=1,
\qquad
\omega = e^\frac{2\pi \imath}{p}.
\end{equation}
 In the standard basis, the matrix $X$ flips the spin cyclically  while the $Z$ acts diagonally:
\begin{equation}
\label{ZXmat}
\langle n' |Z|n\rangle=\omega^{n-1}\delta_{n'n},
\qquad
\langle n'| X|n\rangle=\delta_{n'-1\, n},
\qquad
n,n'=1,\dots,p,
\end{equation}
where  the difference in the Kronecker delta is taken modulo $p$.
The extended Pauli operators are unitary but not Hermitian except for $p=2$ case:
\begin{equation}
\label{Xunitary}
X^+X=Z^+Z=1,
\qquad
X^+=X^{p-1}, \qquad Z^+=Z^{p-1}.
\end{equation}

\medskip

\paragraph*{$\mathbb{Z}_p\times \mathbb{Z}_p$ invariance.}
From the integrability viewpoint, the cluster Hamiltonian \eqref{H} has a quite simple structure:
it is composed of mutually commuting local stabilizers,
which are unitary equivalent to the onsite spin flip operators $X_i$ producing a trivial, $p$-state
Potts model.
Both systems possess a special symmetry, which reveals their topological properties.
This symmetry is formed by simultaneous cyclic permutations of all spins located separately on  the
even and odd positions:
\begin{gather}
\label{comHXeo}
[H,X_\text{even}]=[H,X_\text{odd}]=[X_\text{even},X_\text{odd}]=0,
\\
\label{Xeo}
X_\text{even}=X_2X_4\ldots X_{2L},
\qquad
X_\text{odd}=X_1X_3\ldots X_{2L-1}.
\end{gather}
The elements $X_\text{odd}$ and $X_\text{even}$ produce the group
$\mathbb{Z}_p\times \mathbb{Z}_p=\mathbb{Z}^\text{odd}_p\times \mathbb{Z}^\text{even}_p$
\eqref{Xunitary}.

\medskip

\paragraph*{Trivial phase: noninteracting Potts model.}
As was mentioned above, the cluster Hamiltonian \eqref{H} is obtained from the standard Potts model by the unitary transformation:
\begin{equation}
\label{Htr}
H=UH_\text{trivial} U^+,
\qquad
H_\text{trivial} =-\frac12\sum_{l=1}^{N} (X_l + X_l^+).
\end{equation}
The latter may be regarded as an evolution operator generated by the alternating
nearest-neighbor Ising (Potts) Hamiltonian in  the diagonal basis, which respects
the $\mathbb{Z}_p\times \mathbb{Z}_p$
symmetry:
\begin{equation}
\label{U}
U|n_1\dots n_N\rangle = \omega^{\sum_{l=1}^{N} (-1)^{l-1}n_ln_{l+1}} |n_1\dots n_N\rangle,
\qquad
[U, X_\text{even}] = [U, X_\text{odd}] = 0.
\end{equation}
The expressions for the local stabilizers, which constitute the cluster model, depend on the site's parity:
\begin{equation}
\label{UXU+}
UX_{2r}U^+ = Z_{2r-1} X_{2r} Z^+_{2r+1},
\qquad
UX_{2r+1}U^+ = Z^+_{2r} X_{2r+1} Z_{2r+2}.
\end{equation}

The operator \eqref{U} has a matrix product structure, rather than representing a disjoint product
of local unitaries and, hence,  is capable of producing a short-range
entanglement when acting on a product state.
Note that in quantum computation,
this entangler defines a quantum circuit, which is composed of successive products of
alternating controlled-$Z$ quantum gates acting on the adjacent sites (qudits):
\begin{equation}
\label{U-CZ}
U=\prod_{\text{odd $l$}}CZ_{l\,l+1}\prod_{\text{even $l$}}CZ_{l\,l+1}^+,
\qquad
CZ_{ij}|n_1\dots n_N\rangle=\omega^{n_in_j}|n_1\dots n_N\rangle.
\end{equation}

The trivial system is in the disordered phase with a unique, gapped ground state, which  is the
uniform superposition of all states in the $Z$-basis:
\begin{equation}
\label{gs-triv}
|0\rangle_\text{trivial} = p^{-\frac{N}2}\sum_{n_1,\dots, n_N} |n_1\dots n_N\rangle.
\end{equation}
It is the  product of the symmetric states $p^{-\frac{1}2}\sum_{n=1}^p |n\rangle$.
The unitary map \eqref{U} entangles the trivial state above
preserving the symmetry.
As a result, the ground state of the shifted Hamiltonian \eqref{H} acquires
additional phase factors in the above superposition giving rise to the decorated domain
wall structure:
\begin{equation}
\label{gs}
|0\rangle = U|0\rangle_\text{trivial} = p^{-\frac{N}2} \sum_{n_1,\dots, n_N}  \omega^{\sum_{l=1}^{N} (-1)^{l-1}n_l n_{l+1}} |n_1\dots n_N\rangle.
\end{equation}

Thus, this state together with its parent Hamiltonian, is in nontrivial topological phase protected by the symmetry
$\mathbb{Z}_p\times \mathbb{Z}_p$.
Moreover, the powers $U^2$, $U^3$, \dots $U^{p-1}$
generate Hamiltonians that exhibit other (nonequivalent) nontrivial SPT phases.
 This is in agreement
with the general classification rule of SPT phases \cite{wen12}.
In one dimension,
they are given by the second cohomology group of the symmetry, which, in fact,
is in one-to-one correspondence with the nonequivalent projective representations \cite{wen11}.
In our case, there are exactly $H^2(\mathbb{Z}_p\times \mathbb{Z}_p,U(1))\equiv \mathbb{Z}_p$   different
phases.

\medskip

\paragraph*{String order parameter.}
The trivial ground state is in ferromagnetic order in $X$-diagonal basis while is disordered in the $Z$-basis
since $X_i|0\rangle_\text{trivial}=1$ while $Z_i|0\rangle_\text{trivial}=0$.
Therefore, the vacuum correlations of the products of Pauli $X$
operators on different sites equal unity.
Meanwhile,  the similar correlations for the $Z$ operators vanish:
\begin{align*}
\langle X_i\rangle_\text{trivial}&=\langle X_iX_j\rangle_\text{trivial}
=\langle X_i X_j X_k\rangle_\text{trivial}=\ldots= 1,
\\
\langle Z_{i}\rangle_\text{trivial}&=\langle Z_{i} Z_{j}\rangle_\text{trivial}=\langle Z_i Z_j Z_k\rangle_\text{trivial}=\ldots=0,
\end{align*}
where the shortened notation $\langle \ldots \rangle:=\langle0| \ldots |0\rangle$ is used for the ground-state
expectation value.

Since the unitary entangler is diagonal in the $Z$ basis, the second equation above remains valid
for the SPT state too.
Among the products of $X$ operators, there are special ones whose behavior under
the same entangler is quite simple. They are formed by
the strings  of $m$ successive  operators located on the even or odd sites, $X_iX_{i+2}\dots X_{i+2m-2}$,
depending on the parity of $i$.
Any such string acquires two additional $Z$ and $Z^+$ operators at the ends,
which follows from Eq.~\eqref{UXU+}.
Thus,  the SPT ground state is characterized by the two (even and odd) string order
parameters defined by the nonzero vacuum expectation value:
\begin{equation}
\label{string}
\langle Z_{2r-1}X_{2r}X_{2r+2}\dots X_{2r+2m-2}Z^+_{2r+2m-1} \rangle =
\langle Z^+_{2r}X_{2r+1}X_{2r+3}\dots X_{2r+2m-1}Z_{2r+2m}\rangle=1 .
\end{equation}
Here, the presence of the intermediate $X$ operators is essential and distinguishes
the string order from the usual ferromagnetic
order. They reveal a hidden topological order in the nontrivial SPT
phase. The string order parameter of this type was first observed in the spin-1
Heisenberg  chain (Haldane chain)
\cite{KT}.

\medskip

\paragraph*{Cluster chain with open boundaries.}
In contrast to the periodic case, the protecting symmetry does not impose
any restriction on the length of the
chain with free boundaries.
The odd and even components of the  cluster Hamiltonian \eqref{H} with open boundary conditions
then acquire the following form:
\begin{align}
\label{Hodd-open}
&H_\text{odd} = -\frac{1}{2}\sum_{r=1}^{L-1}  Z_{2r}^+X_{2r+1}Z_{2r+2} + \text{H. c.},
\\
\label{Heven-open}
&H_\text{even} = -\frac{1}{2}\sum_{r=1}^{L'} Z_{2r-1}X_{2r}Z_{2r+1}^+  + \text{H. c.}
\end{align}
Here the  second Hamiltonian's length depends on the parity of the total size:
\begin{equation}
\label{L'}
L'=\begin{cases}
L-1 & \text{for  chain with $N=2L$ sites},
\\
L & \text{for  chain with $N=2L+1$ sites}.
\end{cases}
\end{equation}
Respectively, the summation limits in both the noninteracting Potts Hamiltonian \eqref{Htr} and
the unitary entangler \eqref{U}
must be corrected:
\begin{equation}
\label{Htr-open}
H_\text{trivial} =-\frac12\sum_{l=2}^{N-1} (X_l + X_l^+),
\qquad
U|n_1\dots n_N\rangle = \omega^{\sum_{l=1}^{N-1} (-1)^ln_l n_{l+1}} |n_1\dots n_N\rangle.
\end{equation}
Thus, the ground state of the cluster model with free boundaries undergoes slight changes compared to the closed case
\eqref{gs}:
\begin{equation}
\label{gs-open}
|0\rangle   =  p^{-\frac{N}2} \sum_{n_1,\dots, n_{N}}  \omega^{\sum_{l=1}^{N-1} (-1)^l n_in_{l+1}} |n_1\dots n_N\rangle.
\end{equation}
Note that in contrast to the periodic case \eqref{gs}, the  boundary spins $n_1$ and $n_N$
may be excluded from the summation procedure above
leaving them as $p^2$ parameters the above state depends on: $|0\rangle \to |0\rangle_{n_1n_N}$.
The reason of this  degeneracy is  the left
$Z_1,X_1$  and right $Z_N,X_N$ boundary operators which
do not contribute to the trivial Hamiltonian \eqref{Htr-open}.
The edge $X$ operators pivot under the unitary entangler but unlike to the bulk case \eqref{UXU+},
they  wear only a single nearby $Z$ operator:
\begin{equation}
\label{edge}
UX_1U^+=X_1Z_2,
\qquad
UX_NU^+ =
  Z_{N-1}^\mp X_N=
\begin{cases}
Z_{N-1}X_N & \text{for even $N$},
\\
Z_{N-1}^+X_N & \text{for odd $N$}.
\end{cases}
\end{equation}

Thus, the cluster model has a boundary symmetry generated by  the left and right edge operators
which can be set, respectively, as follows:
\begin{equation}
\label{LR}
L_\text{even}=Z^+_1,
\quad
L_\text{odd}=X_1 Z_2,
\qquad
\text{and}
\qquad
R_\text{even}=Z_{N-1}^\mp X_{N},
\quad
R_\text{odd}= Z^+_{N}.
\end{equation}
%
%
%
They are responsible for the complete spontaneous  symmetry breaking for open
boundaries. Note that the odd part of this  symmetry \eqref{Xeo}
is now generated by the element
$X_\text{odd}=X_1X_3\ldots X_{2L'-1}$.
Moreover, when applied to the ground state, the $\mathbb{Z}_p\times \mathbb{Z}_p $  generators
are reduced to the edge symmetries:
\begin{equation}
\label{Xgs}
X_\text{odd}|0\rangle=L_\text{odd} R_\text{odd} |0\rangle,
\qquad
X_\text{even}|0\rangle=L_\text{even} R_\text{even} |0\rangle,
\end{equation}
which can be verified using the stabilizer structure of the cluster model.
All  states of the $p^2$ fold  degenerate vacuum  are obtained from  the
$X_\text{odd}^nX_\text{even}^m$ monomial action
to the ground state
with
$0\le n,m \le p-1$.

The left  (right)  generators \eqref{LR}, when considered separately, no longer commute.
Instead, they obey the generalized Pauli algebra \eqref{ZX}
which can be regarded as a projective representation of the $\mathbb{Z}_p\times \mathbb{Z}_p$ symmetry
that protects topological phases.
This phenomenon is known as the symmetry fractionization.
Its equivalence class is given by the aforementioned second cohomology coinciding with a single  $\mathbb{Z}_p$ group
whose elements  characterize different SPT phases.

Note that for the two-dimensional cluster model defined on triangular lattice,
the gapless excitations live inside the boundary spin chain which has  more complex
structure than the  bulk system. Its local terms no longer commute with each other  \cite{LG}.
In the Potts case even numerical studies are required
in order to reveal the low-energy behavior \cite{my23}.



\section{Inhomogeneous parafermion cluster chain}
\label{sec:para}

\paragraph*{Lattice parafermions.}
Parafermions are anyons that obey fractional exchange statistics (parastatistics),
which lie between those of bosons and fermions.
They generalize Majorana fermions and, like anyons, exist only in one and two dimensions.
Lattice parafermions are described by operators that satisfy the following algebra:
 \begin{equation}
 \begin{aligned}
 \label{para}
 \chi_i^+\chi_i=1,
 \qquad
\chi_i^p=1,
\qquad
\chi_i\chi_j = \omega \chi_j\chi_i
\qquad
\text{with}
\quad
i<j,
\end{aligned}
\end{equation}
where the fractional exchange phase factor $\omega$ is defined earlier \eqref{ZX}
for an integer $p\ge 2$.

For $p=2$, we use the conventional notation for particle operators: $\chi=\gamma$.
In this case, a particle coincides with its own quasiparticle,
$\gamma=\gamma^+$, the exchange statistics is fermionic, $\omega=-1$,
and the parafermions reduce to Majorana (real) fermions. Two such fermions
constitute a single Dirac (complex) fermion, described by creation-annihilation
operators: $c^\pm=\frac12(\gamma_1\pm \imath\gamma_2)$.
A similar description exists for more general fractional values of  the
parameter $\omega$, and the related particles are called Fock parafermions \cite{fock-par,zero-modes, xu17}.
However, for $p >2$ the construction becomes polynomial and more complicated.
The  exchange relation between a particle
and antiparticle follows from the standard commutation rules for parafermions \eqref{para}:
\begin{equation}
\label{para+}
\chi_i\chi^+_j=\omega^{-1} \chi^+_j\chi_i,
\qquad
\chi_i^+\chi_j=\omega^{-1} \chi^+_j\chi_i,
\qquad
i<j.
\end{equation}

\medskip

\paragraph*{Fradkin-Kadanoff transformation.}
The lattice parafermions may be expressed in terms of the onsite spins, represented in terms
of the generalized Pauli  matrices \eqref{ZX},
by the Fradkin-Kadanoff transformation which
extends the well-known
Jordan-Wigner  transformation to the particles with fractional statistics \cite{FK}:
\begin{align}
\label{FK}
\chi_{2l-1}= Z_l\prod_{k<l}X_k,
\qquad
\chi_{2l}
=\epsilon\chi_{2l-1} X_l
= Y_l\prod_{k< l}X_k.
\end{align}
Therefore, there are  two different parafermions associated with each lattice site,
and the total number of these particles equals $2N$.
Here, a $p$-dimensional analog of the Pauli matrix $\sigma_y$
is introduced:
\begin{equation}
\label{Y}
Y=\epsilon ZX, \qquad
\epsilon = \omega^\frac{p-1}{2} = -\omega^{-\frac12}.
 \end{equation}
It constitutes the following algebra together with the previously defined $X,Y$ operators \eqref{ZX}:
$$
ZY=\omega YZ,
\qquad
YX=\omega XY,
\qquad
Y^p=1.
$$

Note that  the Fradkin-Kadanoff transformation \eqref{FK}  preserves the unitarity: all operators presented are unitary.
The complexity increases as the index value grows.
The first few parafermions have the simplest expressions:
\begin{equation}
\label{chi1-4}
\chi_1=Z_1,
\qquad
\chi_2=Y_1,
\qquad
\chi_3=X_1 Z_2,
\qquad
\chi_4=X_1 Y_2.
\end{equation}

The inverse transformation can also be derived. As a result, the lattice spins
  are re-expressed in terms of the parafermions as follows:
\begin{align}
X_l&=\epsilon^{-1}\chi_{2l-1}^+\chi_{2l},
\label{Xl}
\\
Z_l&=\epsilon^{l-1}\chi_{2l-1}\chi^+_{2l-2}\chi_{2l-3}\chi_{2l-4}^+\ldots  \chi_2^+\chi_1,
\label{Zl}
\\
Y_l&=\epsilon^{l-1}\chi_{2l}\chi^+_{2l-2}\chi_{2l-3}\chi_{2l-4}^+\ldots  \chi_2^+\chi_1.
\label{Yl}
\end{align}
 The simplest representation has the element $X_l$ which is proportional to the parity operator
 of two neighboring parafermions \cite{nepal,hong,loss}. The $Z_l$ is related to the total
 parity of $2l-1$  successive particles  starting from the first.

\medskip

\paragraph*{$\mathbb{Z}_p\times \mathbb{Z}_p $ symmetry generators via parafermions.}
The first formula in the above series  immediately provides a parafermionic representation for the symmetry
generators of the cluster model \eqref{Xeo}:
\begin{equation}
\label{Xeo-par}
\begin{aligned}
X_\text{even}&=\epsilon^{-L} \chi_3^+\chi_4\chi_7^+\chi_8\ldots \chi_{4L-1}^+\chi_{4L},
\\
X_\text{odd}&=\epsilon^{-L'} \chi_1^+\chi_2\chi_5^+\chi_6\ldots \chi_{4L'-1}^+\chi_{4L'},
\end{aligned}
\end{equation}
where the $L'$ depends on the length parity of the spin chain \eqref{L'}.
Their combination generates the diagonal $\mathbb{Z}_p$ subgroup
expressed as an altering product over all parafermions:
\begin{equation}
\label{chi-last}
\chi_{2N+1}:=X_\text{odd}X_\text{even}=\epsilon^{-N} \chi_1^+\chi_2\chi_3^+\chi_4\ldots \chi_{2N-1}^+\chi_{2N}.
\end{equation}
It obeys a simple commutation relation with parafermions, easily verifable using the  algebra \eqref{par}:
\begin{equation}
\label{Xall-chi}
\chi_i \chi_{2N+1}=\omega \chi_{2N+1}\chi_i, \qquad 1\le i\le 2N.
\end{equation}
The element \eqref{chi-last}, which describes the total parafermion parity,  resembles a new composite parafermion added to the system.
Therefore, it can be included in  the parafermion algebra \eqref{para}
as the last member:
\begin{align}
\label{para2}
&\chi_i^p=1,
\qquad
\chi_i\chi_j = \omega \chi_j\chi_i
\qquad
\text{where}
\qquad
1\le i<j\le 2N+1.
\end{align}
In the Ising case when  $p=2$,  the particles reduce to the Majorana fermions being
equivalent to the $2N$ Euclidean  gamma matrices. Then the introduced element
\eqref{chi-last} will also be real and obey the same anticommutation relations with
all primary  gamma matrices, playing the role of the $\gamma^5$ in the four-dimensional Minkowski space.

It is worth noting that the inverse Fradkin-Kadanoff  representations for $Z_l$ and $Y_l$ spin
operators may be written in terms of the composite parafermion $\chi^+_{2N+1}$.
Then the string of the particle operators starts from the end but not from the beginning.
It is easy to check that the following relation holds:
\begin{equation}
\label{Zl'}
Z_l=\epsilon^{l-N-1}\chi_{2l}\chi^+_{2l+1}\chi_{2l+2}\dots\chi^+_{2N-1}\chi_{2N}\chi^+_{2N+1}.
\end{equation}
This representation is particularly convenient for the spin-$Z$ operators close to the end of the chain.
For the last two spins, it gives the simplest expressions:
\begin{equation}
\label{ZN}
Z_N= \epsilon^{-1}\chi_{2N}\chi^+_{2N+1},
\qquad
Z_{N-1}= \omega\chi_{2N-2}\chi^+_{2N-1}\chi_{2N}\chi^+_{2N+1}.
\end{equation}

\medskip

\paragraph*{Spin representations of parafermion bilinears.}
The Fradkin-Kadanoff transformation \eqref{FK}  is essentially nonlocal and may affect the topological
properties of the system.  Moreover,  the bilinear operators of the form
$\chi^+_i\chi_j$ with $i<j$,  being expressed in terms of the spin operators,  contain the product of
all intermediate spin-$X$ matrices between the two sites equipped by other matrices. In particular,
for even $i$ and odd $j$ the following spin representation for the aforementioned bilinear expression   is valid:
\begin{equation}
\label{bil-zz}
\chi_{2l}^+\chi_{2(l+m)-1}= \epsilon Z_l^+ X_{l+1}X_{l+2}\dots X_{l+m-1} Z_{l+m}.
\end{equation}
For the remaining values of $i,j$ indices,
the string is bounded by other Pauli matrices:
\begin{equation}
\label{bil}
\begin{aligned}
\chi_{2l}^+\chi_{2(l+m)}&=\epsilon Z_l^+X_{l+1}X_{l+2}\dots X_{l+m-1}Y_{l+m},
\\
\chi_{2l-1}^+\chi_{2(l+m)-1}&=\epsilon W_lX_{l+1}X_{l+2}\dots X_{l+m-1}Z_{l+m},
\\
\chi_{2l-1}^+\chi_{2(l+m)}&= \epsilon W_l X_{l+1}X_{l+2}\dots X_{l+m-1}Y_{l+m}.
\end{aligned}
\end{equation}
Here we have introduced an onsite spin operator
\begin{equation}
\label{W}
W=\epsilon^{-1}Z^+X,
\end{equation}
which is a time-reversal version of the second generalized Pauli
matrix (see Eq.~\eqref{TR} below).
It is easy to see that it obeys the following
relations:
\begin{equation}
\label{ZW}
W^p=1,
\qquad
ZW=\omega WZ,
\qquad
XW=\omega WX,
\qquad
YW=\omega^2 WY=\omega X^2.
\end{equation}
Note that in $p=2$ case with the usual Pauli matrixes,
the new definition does not make    sense since $W=-Y$. For $p\ge3$,
the four Pauli matrices  can be considered together \cite{tsui}.
With the applied notations in Eqs.~\eqref{bil},  each lattice site in the string is
represented by a single spin operator.
In all equations \eqref{bil-zz}, \eqref{bil},  the restriction $ m\ge 1$ is supposed except for the
last equation, which holds also for the $ m= 0$ case when it reduces to the parafermion representation of the
  $X_l$ \eqref{Xl}.
Note that the quantities \eqref{bil-zz}, \eqref{bil} differ from the string order operators
of the Potts-based  cluster model, in which the product of the $X$ matrices is taken over the lattice sites
 with a specific parity
 \eqref{string}.

Consider  now a few simplest cases of above bilinear operators. Note that for $m=1$,
the intermediate matrices are absent in Eqs.~\eqref{bil}, \eqref{bil-zz}, and we obtain:
\begin{align}
\label{m=1-zz}
\chi_{2l}^+\chi_{2l+1}&= \epsilon Z^+_l Z_{l+1},
\\
\label{m=1-zy}
\chi_{2l}^+\chi_{2l+2}&= \epsilon Z^+_{l}Y_{l+1} =\omega^{-1}Z^+_lZ_{l+1}X_{l+1},
\\
\label{m=1-wz}
\chi_{2l-1}^+\chi_{2l+1}&=\epsilon W_l Z_{l+1} =  Z^+_lZ_{l+1}X_l,
\\
\label{m=1-wy}
\chi_{2l-1}^+\chi_{2l+2}&= \epsilon W_l Y_{l+1} = \epsilon Z^+_l Z_{l+1} X_l X_{l+1}.
\end{align}
%
Next, for  $m=2$, the spin representations contain a single intermediate $X$ operator, and we arrive at the
following expressions:
\begin{align}
\label{m=2-zz}
\chi_{2l-2}^+\chi_{2l+1}& = \epsilon Z^+_{l-1} X_{l}Z_{l+1},
\\
\label{m=2-zy}
\chi_{2l-2}^+\chi_{2l+2} &=  \epsilon Z^+_{l-1}  X_{l}Y_{l+1} = \omega^{-1}  Z^+_{l-1} Z_{l+1} X_{l}X_{l+1},
\\
\label{m=2-wz}
\chi_{2l-3}^+\chi_{2l+1}& = \epsilon W_{l-1}  X_{l}Z_{l+1} = Z^+_{l-1}Z_{l+1}X_{l-1}X_{l},
\\
\label{m=2-wy}
\chi_{2l-3}^+\chi_{2l+2} &=\epsilon W_{l-1} X_{l} Y_{l+1} = \epsilon  Z^+_{l-1} Z_{l+1} X_{l-1} X_{l}X_{l+1}
\end{align}
where the index substitution  $l\to l-1$ is applied in order to get more symmetric expressions.

\medskip

\paragraph*{Parafermion representation of Potts-based cluster model with open boundaries.}
Now let us focus on the parafermionic representation of
the cluster Hamiltonian \eqref{H}. In this section, we consider  the free system \eqref{Hodd-open},
\eqref{Heven-open}, \eqref{L'}. Other cases will be treated
below in this article.

First note that Eq.~\eqref{m=2-zz} encodes the terms from
the odd part of the model \eqref{Hodd-open}
in terms of the parafermions:
\begin{equation}
\label{z+xz}
 Z^+_{l-1} X_lZ_{l+1}=\epsilon^{-1}\chi_{2l-2}^+\chi_{2l+1}.
\end{equation}
Then, using also  Eq.~\eqref{Xl}
and the particle exchange relations \eqref{para}, \eqref{para+},
one can write
the parafermionic representation  for the even
stabilizers  in the Hamiltonian \eqref{Heven-open}:
\begin{equation}
  \label{zxz+}
 Z_{l-1} X_lZ^+_{l+1}=\left( Z^+_{l-1} X_lZ_{l+1}\right)^+ X_l^2
 =\epsilon^{-1}\chi_{2l+1}^+\chi_{2l-2}\left(\chi_{2l-1}^+\chi_{2l}\right)^2
=\epsilon^{-5}\chi_{2l-2}\chi_{2l-1}^{+2}\chi_{2l}^2\chi_{2l+1}^+.
\end{equation}

As a result, the generalized cluster Hamiltonian \eqref{H} with open boundary conditions \eqref{Hodd-open}
and an even number of sites ($N=2L$),
being expressed in terms of the parafermions,
acquires the following structure:
\begin{equation}
\label{Hpar1}
H=-\frac{1}{2\epsilon}
\sum_{r=1}^{L-1} \left(
\chi_{4r}^+\chi_{4r+3}
+
\omega^2\chi_{4r-2}\chi_{4r-1}^{+2}\chi_{4r}^{2}\chi_{4r+1}^+
\right)
+\text{H. c.}
\end{equation}
It consists of two commuting Hamiltonians, described by the first and second
local terms, respectively.
The same model  defined on an odd-length chain  ($N=2L+1$) contains two more terms of the second type,
namely the monomial
\begin{equation}
\label{term-per}
-\epsilon^{-5}\chi_{4L+1}\chi_{4L+3}^{+2}\chi_{4L+4}^{2}\chi_{4L+5}^+
\end{equation}
and its Hermitian conjugate.

A straightforward application of the generalized Jordan-Wigner transform
complicates the physical interpretation of the resulting system \eqref{Hpar1}.
Indeed, the first part of the model is bilinear in parafermions and may be described
as free particles. However,  in general, the second part differs drastically.
It   consists of sixth-order
monomials, which may be regarded as a $(\chi^+\chi)^3$-type interaction among
 particles located at the four neighboring positions.
Recall that in the
original spin system, both components contain three-spin interactions among the three
neighboring sites, which only slightly differ from each other \eqref{Hodd}, \eqref{Heven}.
Let us consider the two particular values of the exchange phase parameter when the
above parafermion system simplifies.

For the $p=2$ case,  the parafermions are reduced to the  Majorana fermions $\gamma_i$:
\begin{equation}
\label{gamma}
\epsilon=\imath,
\qquad
\gamma_i=\chi_i:
\qquad
\{\gamma_i,\gamma_j\}=2\delta_{ij},
\end{equation}
and the Hamiltonian \eqref{Hpar1} simplifies  drastically.
The particle operators become self-conjugate, $\gamma^+=\gamma$, and their squares in the interaction terms disappear.
As a result, the even subsystem  also becomes a  bilinear combination of particle operators, and the
entire Hamiltonian   reduces to two decoupled  Kitaev chains:
 \begin{equation}
H=\imath
\sum_{l=1}^{N-1}
\gamma_{2l}\gamma_{2l+3}.
\end{equation}
The above system coincides with the $\alpha=2$ chain in the classification scheme applied to the
fermionic representation of the generalized cluster models  \cite{verresen}.

In the case of the  $p=3$ state parafermions, the fractional exchange phase takes the value
$\epsilon=\omega=e^\frac{2\pi \imath}{3}$, and, due to the relation $\chi^2=\chi^+$,
the members of the $H_\text{even}$ are reduced to a four-particle interaction of the  type $(\chi^+\chi)^2$.
Thus, the Hamiltonian  simplifies  slightly \cite{xu17}:
\begin{equation}
\label{Hmaj}
H=-\frac{1}{2}
\sum_{r=1}^{L-1} \left(
\omega^{-1}\chi_{4r}^+\chi_{4r+3}
+
\omega\chi_{4r-2}\chi_{4r-1}\chi_{4r}^+\chi_{4r+1}^+
\right)
+\text{H. c.}
\end{equation}
Note that a  four-particle interaction term of a similar type arose recently
in the extended parafermion chain \cite{ext}.

In the next section,  the time-reversal particle operators will be used in order to present the
parafermionic system \eqref{Hpar1} in a bilinear form similar to the Majorana-fermion chain
\eqref{Hmaj}, which will be more amenable for revealing the symmetries and edge zero-energy modes.

\medskip


\section{Splitting into two similar chains with standard and time reversal parafermions: open boundaries}
\label{sec:native}
In this section,  we derive an alternative bilinear representation of the $p$-state open cluster
model which involves the time-reversal parafermions. The symmetries and
parafermionic edge modes, which elicit the nontrivial SPT phase and are responsible
for the ground-state degeneracy, are studied.
First, we explore  the behavior of particle operators under time-reversal and lattice parity transformations.

\medskip

\paragraph*{Time reversal.}
For spinless particles, the time-reversal transformation
is reduced to the complex conjugation of the wave function $\psi$. It is the simplest antiunitary involutive operation:
$$
\mathcal{T}\psi=\psi^*,
\qquad
\mathcal{T}^2=1.
$$
Under time reversal,  the extended Pauli matrices \eqref{ZX}, \eqref{Y}, and \eqref{W} map to each other
\cite{batchelor}:
\begin{equation}
\label{TR}
\mathcal{T} X\mathcal{T}^{-1}=X,
\qquad
\mathcal{T} Z\mathcal{T}^{-1}=Z^+,
\qquad
\mathcal{T} Y \mathcal{T}^{-1}=W.
\end{equation}
Therefore, the $\mathbb{Z}_p\times \mathbb{Z}_p$  group  generated by the elements
 \eqref{Xeo}  is $\mathcal{T}$-invariant.

In the following, we will use the barred symbols (instead of the asterisk superscript) in order to denote the complex conjugate  operators:
 \begin{equation}
 \label{time-chi}
 \bar\chi:=\chi^*=\mathcal{T}\chi\mathcal{T}^{-1}.
 \end{equation}
The barred parafermions obey the inverse statistics (with the fractional phase factor $\omega$ replaced by $\omega^{-1}$) compared with
 the initial one \eqref{para} \cite{TR}:
 \begin{equation}
 \label{para'}
\bar\chi^p_i=1,
\qquad
\bar\chi_i\bar\chi_j = \omega^{-1} \bar\chi_j\bar\chi_i,
\qquad
i<j.
\end{equation}
Using Eqs.~\eqref{TR}, the time-reversal version of the Fradkin-Kadanoff transformation \eqref{FK} is easily obtained:
\begin{align}
\label{FK'}
\bar\chi_{2l-1}= Z^+_l\prod_{k<l}X_k,
\qquad
\bar\chi_{2l}
= W_l\prod_{k< l}X_k.
\end{align}
The corresponding bilinear operators are just the complex conjugates of the corresponding expressions \eqref{bil-zz}, \eqref{bil}:
\begin{equation}
\label{bilin'}
\begin{aligned}
\bar\chi_{2l}^+\bar\chi_{2(l+m)-1}&= \epsilon^{-1} Z_l X _{l+1}\dots X _{l+m-1} Z^+_{l+m},
\\
\bar\chi_{2l}^+\bar\chi_{2(l+m)}&=\epsilon^{-1} Z_l X _{l+1}\dots X _{l+m-1}W_{l+m},
\\
\bar\chi_{2l-1}^+\bar\chi_{2(l+m)-1}&=\epsilon^{-1} Y_lX_{l+1}\dots X_{l+m-1}Z^+_{l+m},
\\
\bar\chi_{2l-1}^+\bar\chi_{2(l+m)}&= \epsilon^{-1} Y_l X _{l+1}\dots X _{l+m-1}W_{l+m},
\end{aligned}
\end{equation}

The mixture of the two sets of parafermions, $\chi$ and $\bar\chi$, obeys a generalized parastatistics with
different values of exchange constant among the particle pairs. Namely, in most cases when  the particle indices
differ, they obey
a simple homogeneous commutation rule:
\begin{equation}
\label{chichi'}
\chi_i\bar\chi_j = \omega \bar\chi_j\chi_i,
\qquad
i\ne j.
\end{equation}
An exception occurs when we exchange the particle with its own time reversal, and the result
depends on the parity of the particle's index:
\begin{equation}
\label{chichi'2}
\chi_{2l}\bar\chi_{2l} = \omega^2 \bar\chi_{2l}\chi_{2l},
\qquad
\chi_{2l-1}\bar\chi_{2l-1} =  \bar\chi_{2l-1}\chi_{2l-1}.
\end{equation}

In the particular  case of the Majorana fermions \eqref{gamma}, things simplify drastically.
Remember that Majoranas are  real particles coinciding with their antiparticles. Half of them alter their sign
under the time reversal while the rests are not changed at all:
\begin{equation}
\label{time-maj}
 \bar\gamma_i= (-1)^{i-1}\gamma_i.
\end{equation}
Note that the above equation together with Eq.~\eqref{chichi'2}
is in agreement with the fact that the Majorana operator is involutive: $\gamma_i^2=1$.

Clearly, the time-reversal particles are not independent, but are expressed in terms of the original ones.
The connection among them can be obtained from the following relations,
which are implied by the Fradkin-Kadanoff transformation:
\begin{equation}
\label{bar-chi-Z}
\bar\chi_{2l-1} = Z_l^{+2} \chi_{2l-1},
\qquad
\bar\chi_{2l} = \omega Z_l^{+2} \chi_{2l}.
\end{equation}
Applying the inverse transformation for the Pauli $Z$ matrices \eqref{Zl} further, one can get  the final expressions for
the time-reversal particle operators.
In particular, the equations \eqref{bar-chi-Z} result in an equivalence among the respective
bilinear terms:
\begin{equation}
\bar\chi^+_{2l-1} \bar\chi_{2l} =\omega\chi_{2l-1}^+ \chi_{2l}.
\end{equation}

Finally, we mention that the $\mathbb{Z}_p\times \mathbb{Z}_p$  group, which produces the SPT phases and
is  generated by the elements
 \eqref{Xeo},  is $\mathcal{T}$-invariant
 because its generators are composed of the $\mathcal{T}$-invariant  operators
 \eqref{TR}.
Hence, the composite parafermion \eqref{chi-last}  remains unchanged as well:
\begin{equation}
\label{bar-chi-last}
\bar\chi_{2N+1} = \chi_{2N+1}.
\end{equation}
\medskip

\paragraph*{Parity transformation.}
The lattice parity transformation flips the lattice nodes with respect to the middle
of the chain. More explicitly, the onsite Pauli matrices transform as follows:
\begin{equation}
\label{par}
\mathcal{P}X_{l}\mathcal{P}=X_{l'},
\qquad
\mathcal{P}Z_{l}\mathcal{P}=Z_{l'},
\qquad
l'=N+1-l,
\end{equation}
where $l'$ denotes the reflected site. Clearly, it is involutive and consistent with the time-reversal
transformation:
\begin{equation}
\label{Psq}
\mathcal{P}^2=1,
\qquad
[\mathcal{T},\mathcal{P}]=0.
\end{equation}
From Eq.~\eqref{par} it is easy to see that the $\mathbb{Z}_p\times \mathbb{Z}_p$
generators remain invariant or are mapped to each other depending on the parity of
the lattice size:
$$
\begin{aligned}
N=2L:  \qquad & \mathcal{P}X_\text{even}\mathcal{P}=X_\text{odd},
\\
N=2L+1:  \qquad & [\mathcal{P},X_\text{even}]=[\mathcal{P},X_\text{odd}]=0.
\end{aligned}
$$
In both cases, the combined symmetry operator \eqref{chi-last}
remains invariant with respect to the
space parity transformation:
\begin{equation}
\label{Pchi-last}
\mathcal{P}\chi_{2N+1}\mathcal{P} = \chi_{2N+1}.
\end{equation}
%
%
The transformation rules for the basic particle operators can be derived
using the Fradkin-Kadanoff mapping \eqref{FK}. Indeed, using the  reflection rules of the
spin operators \eqref{par}, their algebra  and  transposes:
$X^\tau = X^{-1}$, $Z^\tau =Z$, one obtains the
following rules:
\begin{equation}
\label{Pchi-eo}
\begin{aligned}
\mathcal{P}\chi_{2l}\mathcal{P}&=\varepsilon Z_{l'}\prod_{k=l'}^{2N}X_k=
\varepsilon Z_{l'}\left(\prod_{k=1}^{l'-1}X^\tau_k\right)\chi_{2N+1}=\epsilon \chi^\tau_{2l'-1} \chi_{2N+1},
\\
\mathcal{P}\chi_{2l-1}\mathcal{P}&=Z_{l'}\prod_{k=l'+1}^{2N}X_k=
\omega^{-1} \left(\prod_{k=1}^{l'}X^\tau_k\right)Z_{l'}\chi_{2N+1}
=\epsilon \chi^\tau_{2l'} \chi_{2N+1}.
\end{aligned}
\end{equation}
Recall that the operator $\chi_{2N+1}=X_1X_2\dots X_N$, which generates the diagonal $\mathbb{Z}_p$ symmetry,
represents also a composite parafermion given by the alternating product
of all particle operators \eqref{chi-last}.
It appears here due to the nonlocal nature of the Fradkin-Kadanoff transformation  \eqref{FK}.
Note that the second equation in \eqref{Pchi-eo} follows from the first one
and vice versa, as it is easy to verify using the algebra \eqref{Psq}.
Finally, notice that the above two equations can be combined into a single formula holds
for odd  and even  values of indices:
\begin{equation}
\label{Pchi}
\mathcal{P}\chi_{i}\mathcal{P}=\epsilon \chi^\tau_{2N+1-i} \chi_{2N+1},
\qquad
1\le i\le 2N.
\end{equation}

Consider the restriction of the relation \eqref{Pchi}  to the Majorana fermions \eqref{gamma}.
They  are described by real  operators, so their transposes coincide with the time reversals:
$\gamma_i^\tau=\bar\gamma_i=(-1)^{i-1}\gamma_i$.
Then, the Majorana time-reversal rule \eqref{time-maj} implies:
\begin{equation}
\label{Pgamma}
\mathcal{P}\gamma_{i}\mathcal{P}=\imath(-1)^{i-1} \gamma_{2N+1-i} \gamma_{2N+1}.
\end{equation}

%

\medskip

\paragraph*{$\mathcal{PT}$ transformation.}
By combining the relations
\eqref{time-chi} and \eqref{Pchi}, one obtains
the joint action of the time-reversal
and  lattice parity transformations on the particle operators:
\begin{equation}
\label{PTchi}
\mathcal{PT}\chi_{i}\mathcal{PT}=\epsilon^{-1}\chi^+_{2N+1-i} \chi_{2N+1},
\end{equation}
where the reality of the composite parafermion \eqref{bar-chi-last} is also taken into account.
Note that the combined parity-time  symmetry holds for a wide range
of models that are not conserved separately under  time reversal and the space reflection.
The $\mathcal{PT}$ invariant systems have garnered  significant interest in physics because
 they have real energy spectra when this symmetry is unbroken,
even if the corresponding Hamiltonians are  not Hermitian \cite{PT}.

Finally, note that the mapping
\eqref{PTchi} applied to  the Majorana fermions takes  the following form:
$$
\mathcal{PT}\gamma_{i}\mathcal{PT}=-\imath\gamma_{2N+1-i} \gamma_{2N+1}.
$$

\medskip

\paragraph*{Hamiltonian via time-reversal parafermions.}
As previously discussed, the parafermionic representation of the cluster Hamiltonian obtained in the previous
section
\eqref{Hpar1}, \eqref{term-per}, \eqref{Hpar1-per},
contains the
sixth-order interaction terms \eqref{zxz+}. They form the second, even
part of the system, $H_\text{even}$, and break a visual equivalence with
its odd counterpart,  $H_\text{odd}$.
The symmetry between the two subsystems can be restored if we use
a simpler and more elegant version of the parafermionic Hamiltonian,
which involves  the time-reversal particle
operators \eqref{time-chi}.
Indeed,  the complex conjugate local terms of the $H_\text{even}$ \eqref{Heven} are similar to the local
terms of the $H_\text{odd}$ \eqref{Hodd}. Their particle representation
is given by the complex conjugate version of Eq.~\eqref{z+xz}:
\begin{equation}
\label{zxzp}
 Z_{l-1} X_lZ^+_{l+1}=\epsilon\bar\chi_{2l-2}^+\bar\chi_{2l+1}.
\end{equation}
It is bilinear in $\mathcal{T}$-transformed particle operators. Thus, the  entire system with open boundaries
can be rewritten  now in terms of the original and barred particle operators as follows:
\begin{equation}
\label{Hpar2}
H=-\frac{1}{2\epsilon}\sum_{r=1}^{L-1} \chi_{4r}^+\chi_{4r+3}  -\frac\epsilon{2}\sum_{r=1}^{L'}\bar\chi_{4r-2}^+\bar\chi_{4r+1}
\,+\,\text{H. c.}
\end{equation}
This  represents an alternative version of the parafermionic system \eqref{Hpar1}.
Recall that the upper limit of the second sum  depends on the length parity of the spin
model \eqref{L'}, and the first and second sums describe,
respectively, the odd \eqref{Hodd} and even \eqref{Heven} parts of the
original spin system.

We conclude that the entire system is split into two  subsystems,
each described by its own particles. Namely, the subsystem with the Hamiltonian
 $H_\text{odd}$ is
characterised by the standard parafermions with the indices $i=0,3$ modulo $4$
while the other with $H_\text{even}$ is expressed in terms of the
time-reversal parafermions with the indices $i=1,2\mod4$.

For comparison, below we also present  the parafermionic expression of  the trivial Hamiltonian \eqref{Htr-open} which
immediately follows from the particle representation of the $X_l$ operator \eqref{Xl}.
It contains only the native particles but not their time reversals:
\begin{equation}
\label{Hpar-triv}
H_\text{trivial}=-\frac{1}{2}\sum_{l=2}^{N-1}\left( \epsilon^{-1}\chi_{2l-1}^+\chi_{2l}  + \epsilon\chi_{2l}^+\chi_{2l-1}  \right).
\end{equation}
Another interesting and simplest system containing only nearest-neighboring terms,
describes the potential  of the $p$-state
Potts model:
\begin{equation}
\label{Hpar-zz}
H_\text{ZZ}=-\frac12\sum_{i=1}^{N-1}\left(Z^+_iZ_{i+1}+Z^+_{i+1}Z_{i}\right)=
-\frac12\sum_{l=1}^{N-1}\left( \epsilon^{-1}\chi_{2l}^+\chi_{2l+1}  + \epsilon\chi_{2l+1}^+\chi_{2l}  \right).
\end{equation}
The combination of two models  \eqref{Hpar-triv}, \eqref{Hpar-zz} with omitted second terms in the
bracket, which violates the hermiticity of the resulting
Hamiltonian, produces the homogeneous version
of the free parafermion model \cite{fendley14}:
$$
H_\text{free}= \epsilon^{-1}\sum_{i=1}^{2N-1} \chi^+_i\chi_{i+1},
$$
which provides the parafermion representation of the Baxter's clock model \cite{clock}.
Its solution  is not as trivial as in the Majorana case. Quite recently, a more general
criterion for spin models described by  free parafermions has been formulated
in terms of graph theory
\cite{elman24}.

%
%

Note that the use of the barred particle operators (instead of the usual ones)
in the Hamiltonian \eqref{Hpar2} is necessary in order to
associate the local terms with stabilizers. In fact, the replacement $\bar\chi_i\to \chi_i$ in the second part
will spoil the commutativity among the local terms. Indeed, using the algebra \eqref{para}, it is easy to verify
that the  "intersecting" local parts do not commute except for the fermionic case when $\omega=-1$:
$$
(\chi^+_i\chi_j)(\chi^+_{i'}\chi_{j'})=\omega^2(\chi^+_{i'}\chi_{j'})(\chi^+_i\chi_j)\qquad \text{if} \qquad i<i'<j<j'.
$$

\medskip
\paragraph*{Splitting into two parafermion chains.}
It is worth noticing that each of the two parts, constituting
the $\mathbb{Z}_p^\text{odd}\times \mathbb{Z}_p^\text{even}$ symmetry,
is composed of the particle operators that form the related subsystem.
Indeed, the generator $X_\text{even}$  from Eq.~\eqref{Xeo-par}  describes the total parafermion
parity of the $\chi$-particles that constitute the first subsystem in  Eq.~\eqref{Hpar2}, described by the Hamiltonian
$H_\text{odd}$.   Similarly,  the element $X_\text{odd}$, which
may also be presented as
\begin{equation}
\label{X0-par}
X_\text{odd}=\epsilon^{L'} \bar\chi_1^+\bar\chi_2\bar\chi_5^+\bar\chi_6\ldots \bar\chi_{4L'-1}^+\bar\chi_{4L'},
\end{equation}
is the total parity of the $\bar\chi$-parafermions forming the second subsystem $H_\text{even}$.
Hence, the $p$-state cluster chain together with the symmetry, which protects its
topological order, splits into two separate parafermion chains with $\mathbb{Z}_p^\text{odd}$
and $\mathbb{Z}_p^\text{even}$ topological order. Similarly to fermion parity, the parafermion parity
is always preserved in physical systems. This property eliminates the symmetry-protection requirement
and endows both subchains with intrinsic topological orders.
In one dimension, these phases have
already been classified \cite{quella13,poll13}.

\medskip
Now, let us investigate the \emph{space-time symmetries} of the parafermionic cluster model \eqref{Hpar2}.
Obviously, the  Hamiltonian \eqref{Hpar2} is not $\mathcal{T}$-invariant but transforms
into an equivalent system given by its conjugate:
\begin{equation}
\label{Hpar2T}
\bar H=\mathcal{T}H\mathcal{T}^{-1}
=-\frac{\epsilon}{2}\sum_{r=1}^{L-1} \bar\chi_{4r}^+\bar\chi_{4r+3}  -\frac1{2\epsilon}\sum_{r=1}^{L'}\chi_{4r-2}^+\chi_{4r+1}
\,+\,\text{H. c.}
\end{equation}
Obviously, its ground state differs from that of the original system
\eqref{gs} except in the Majorana case when both models are identical:
\begin{equation}
\label{Tgs}
\bar{|0\rangle}=\mathcal{T} |0\rangle= p^{-\frac{N}2} \sum_{n_1,\dots, n_N}  \omega^{\sum_{l=1}^{N-1} (-1)^{l}n_l n_{l+1}} |n_1\dots n_N\rangle
\end{equation}
with the omitted term $n_Nn_1$ in the exponent  due to the free boundaries.
%
%
%
%
%
%

\medskip

\paragraph*{$\mathcal{P}$ or $\mathcal{PT}$ invariance of the Hamiltonian.}

The behavior of the cluster model under the lattice reflection depends on the length parity of the
spin chain.
For \emph{even} $N$, this transformation maps between local terms of the even and odd Hamiltonians making the
entire system reflection-invariant as it follows from the relations \eqref{par}, or \eqref{Pchi}. It is also easy to see
 that the ground state \eqref{gs} remains invariant and possesses the even parity quantum number:
\begin{align}
\label{PH1}
N=2L:\qquad
&\mathcal{P}H_\text{even}\mathcal{P}=H_\text{odd},
\qquad
\mathcal{P}H\mathcal{P}=H,
\\[8pt]
\label{PH2}
& \mathcal{P}|0\rangle = p^{-\frac{N}2} \sum_{n_1,\dots, n_N}
\omega^{\sum_{l=1}^{N-1} (-1)^{l-1}n_{N-l+1} n_{N-l}} |n_1 n_2\dots n_N\rangle
=|0\rangle.
\end{align}

The same equations imply that for the \emph{odd-length} chain, any local term
of the odd Hamiltonian is reflected into a complex-conjugate term within the same Hamiltonian.
The even Hamiltonian also satisfies this property.
Therefore, both subsystems, together with the entire system, remain invariant under the
simultaneous  application of the parity and time-reversal transformations. Meanwhile,
it is easy to see that for odd $N$, the reflection, like the time reversal \eqref{Tgs}
alters the sign in the coefficient's exponent in the ground-state expansion
\eqref{gs}. Thus, both transformations act identically leaving the ground state
 $\mathcal{PT}$-invariant with the unit quantum number:
\begin{align}
\label{PTH1}
N=2L+1:
\qquad
&[\mathcal{PT},H_\text{even}]=[\mathcal{PT},H_\text{odd}]=[\mathcal{PT},H]=0,
\\[3pt]
\label{PTH2}
& \mathcal{P}|0\rangle = \bar{|0\rangle}, \qquad\mathcal{PT}|0\rangle
=|0\rangle.
\end{align}

\medskip
\paragraph*{Boundary parafermions.}
Consider now the parafermionic modes that appear at the edges of the open
chain \eqref{Hpar2}. Note that the left  $\chi_1$, $\chi_3$ and the right   $\chi_{2N-2}$,  $\chi_{2N}$
boundary operators together with their time reversal counterparts are absent in the Hamiltonian which, in fact,
  does not imply their conservation.

First, consider the \emph{even-length} model which is $\mathcal{P}$-invariant \eqref{PH1}.
Using the parafermionic
algebra \eqref{para}, \eqref{para'}, \eqref{chichi'},  it is easy to verify
that the following three parafermions are conserved:
\begin{equation}
\label{comHb-even}
[H,\chi_1]=[H,\chi_3]=[H,\chi_{2N}]=0.
\end{equation}
Then, their reflected versions, which are proportional to the bilinear operators
$\chi^\tau_k\chi_{2N+1}$ with $k=2N,\,2N-2,\,1$ \eqref{Pchi} listed in the same order,
are also edge invariants.
As a result, the $\mathcal{T}$-counterparts of the edge  particles
also give rise to zero-energy modes:
\begin{equation}
\label{comHb'}
[H,\bar\chi_{2N}]=[H,\bar\chi_{2N-2}]=[H,\bar\chi_1]=0. 
\end{equation}
Meanwhile,  the operators $\bar\chi_3$ and $\chi_{2N-2}$ are not preserved as is easy
to verify.
Using the parafermionic representations of the $Z_1$
\eqref{chi1-4} and
$Z_N$ \eqref{ZN} spin operators,
it is easy to see that the first and last particles are interrelated with their barred versions:
\begin{equation}
\label{bchi*chi}
\bar\chi_1\chi_1 = 1,
\qquad
\bar\chi_{2N}\chi_{2N}=w^{-1}\chi_{2N+1}^2.
\end{equation}
Eventually,   only four parafermions  out of above six ones
remain independent.
As an independent set of edge operators, one can choose, for instance,  the system
consisting of the following single particles:
\begin{equation}
\label{b1}
\beta_1=\chi_1, \qquad \beta_2=\chi_3, \qquad   \beta_3=\chi_{2N},
\qquad
\beta_4=\bar\chi_{2N-2},
\end{equation}
The above list is regrouped  in a way that obeys the conventional arrangement for the parafermions \eqref{para}
with the commutation rules \eqref{chichi'}:
\begin{equation}
\label{alpha}
\beta_i^p=1,
\qquad
\beta_i\beta_j=\omega \beta_j\beta_i,
\qquad
1\le i<j \le 4.
\end{equation}

They form parafermionic versions of the boundary spins modes
appearing due to the fractionalization of the $\mathbb{Z}_p\times\mathbb{Z}_p $
symmetry on the left and right parts
\eqref{Xeo}. These spin modes are also related to each other by the space parity
map as is easy to see from the relations \eqref{edge}.
They can be reexpressed in terms of the edge
parafermions:
\begin{equation}
\label{LR-even}
L_\text{even}=\beta_1^+,
\qquad
L_\text{odd}=\beta_2,
\qquad
R_\text{even}=\mathcal{P}\beta_2\mathcal{P}=\epsilon \beta_4^+\chi_{2N+1},
\qquad*
R_\text{odd}=\mathcal{P}\beta^+_1\mathcal{P}
=\epsilon^{-1}\beta_3^+\chi_{2N+1}.
\end{equation}
In the derivation of the last equation, the second relation in \eqref{bchi*chi} has been  applied.
Before moving to the next case, let us note that the absent edge modes $\beta_2$
and $\beta_3$ belong to the first subsystem whereas the  $\beta_1,\,\beta_4$
 --- to the second one \eqref{Hpar2}. This choice is suggested by the index
 values in  definition \eqref{b1}. [See also below for the translationally invariant case \eqref{Hpar-per-2}.]
From this perspective, a more natural notation for the edge modes can be represented as:
\begin{equation}
\label{b1'}
\beta'_1=\chi_3, \qquad   \beta'_2=\chi_{2N},
\qquad
\beta'_3=\bar\chi_{2N-2}, \qquad \beta'_4=\bar\chi_1,
\end{equation}
while maintaining the standard parafermion ordering \eqref{alpha}.

\smallskip

Now, turn to the \emph{odd-length} chains: $N=2L+1$. Evidently, the left edge parafermions
$\chi_1$ and $\chi_3$ commute with the Hamiltonian. Due to $\mathcal{PT}$
symmetry, their images under this map are also conserved. Therefore, from Eq.~\eqref{PTchi}
it is clear that the following boundary particles are invariants  of the system:
\begin{equation}
\label{comHb-odd}
[H,\chi_1]=[H,\chi_3]=[H,\chi_{2N-2}]=[H,\chi_{2N}]=0. 
\end{equation}
In contrast to the even-length cluster model, there is no time-reversal particle $\bar\chi_i$ among the  invariant edge
modes: all modes are described by operators $\chi_i$. Hence, one can number them by retaining
their original order:
\begin{equation}
\label{b2}
\beta_1=\chi_1, \qquad \beta_2=\chi_3, \qquad   \beta_3=\chi_{2N-2},
\qquad
\beta_4=\chi_{2N}.
\end{equation}

The particle representations of the left and right edge symmetries  \eqref{LR} are provided, respectively, by the
first two equations in \eqref{LR-even} and the relations:
$$
R_\text{even}=\mathcal{PT}\beta_2\mathcal{PT}=\epsilon^{-1} \beta^+_3\chi_{2N+1},
\qquad
R_\text{odd}=\mathcal{PT}\beta^+_1\mathcal{PT}=\epsilon^{-1}\beta_4\chi_{2N+1}^+.
$$
Notice that in contrast to the even $N$ case \eqref{b1'}, the edge parafermion $\chi_{2N-2}=\chi_{4L}$ is associated with the odd part of the Hamiltonian
while the particle $\chi_{2N}=\chi_{4L+2}$ is related to the even one. In this context,
the following designation of the boundary modes might actually make more sense than the above one \eqref{b2}:
\begin{equation}
\label{b2'}
\beta'_1=\chi_3, \qquad   \beta'_2=\chi_{2N-2},
\qquad
\beta'_3=\bar\chi_{2N}, \qquad \beta'_4=\bar\chi_1.
\end{equation}

Thus, one might conclude that the open chain possesses four invariant edge
parafermionic modes \eqref{b1'} or \eqref{b2'}, equally distributed among the two parts: two of them
are located at the boundaries of the odd subsystem while the remaining two --- at the
boundaries of the second, even subsystem.

In the Majorana case, the  boundary invariants are described by the fermions $\gamma_1$, $\gamma_3$,
$\gamma_{2N-2}$,
and $\gamma_{2N}$ irrespective of the boundary condition. The arrangement does not matter due to
the anticommutation relations.

Note that the trivial model \eqref{Hpar-triv} does not contain the four edge particles: $\chi_1$, $\chi_2$, $\chi_{N-1}$, and $\chi_N$
which, as is easy to see, commute with the Hamiltonian. In  the nontrivial phase, the second and third particles listed above are replaced by others,
which manifests a crucial difference between two systems.

\medskip

\section{Splitting into two similar chains: The translationally invariant case.}

\paragraph*{Translation of the spin lattice.}
\label{sec:Tinv}
The translation operator cyclically shifts a multispin state to the right by a single lattice step:
\begin{equation}
\label{T}
T|n_1n_2\dots n_N\rangle=|n_Nn_1\dots n_{N-1}\rangle,
\qquad
TX_iT^{-1}=X_{i\,\text{mod}\, N +1}.
\end{equation}
It produces the $\mathbb{Z}_N$
cyclic group since  $T^N=1$.
Clearly, the translation is compatible with the time reversal and
produces the dihedral group $\mathbb{D}_{N}$ together with the reflection
$\mathcal{P}$.

Clearly, the translation maps the  $\mathbb{Z}_p\times \mathbb{Z}_p$ symmetry
generators to each other but leaves invariant their combination:
\begin{equation}
\label{TX}
T X_\text{even}T^{-1}=X_\text{odd},
\qquad
 T X_\text{odd} T^{-1}=X_\text{even},
 \qquad
T \chi_{2N+1}T^{-1} =  \chi_{2N+1}.
\end{equation}

Let us describe how the translation acts on the primary parafermions. Using their representation via
the spin operators \eqref{FK} together with the inverse relations \eqref{chi1-4}, it is easy to
derive the translation rule for $l<N$:
\begin{align}
\label{Tchi}
T\chi_iT^{-1}=\epsilon^{-1}\chi_1\chi^+_2\chi_{i+2},
\qquad
\text{for} \quad 1\le i \le 2N-2.
\end{align}
Due to the nonlocal nature of the Fradkin-Kadanoff mapping,  the translation rules are more complicated
for the last two particles. In that case, the corresponding expressions contain, in addition, the
composite parafermion operator \eqref{chi-last}:
\begin{equation}
\label{Tchi-N}
\begin{aligned}
T\chi_{2N-1}T^{-1}=\epsilon\chi_1\chi^+_2\chi_1 \chi_{2N+1},
\qquad
T\chi_{2N}T^{-1}=\epsilon\chi_1 \chi_{2N+1}.
\end{aligned}
\end{equation}
More explicitly, the above two equations contain
the additional right factor $\omega^{-1}\chi_{2N+1}$ comparing
with the general transformation rules \eqref{Tchi}.

It is interesting to consider the conversion  rule of the bilinear operators under the cyclic
shift on a single spin node.
It can be derived using Eqs.~\eqref{Tchi}, \eqref{Tchi-N}. Applying  the exchange algebra of parafermions,
the result can be presented in a compact form:
\begin{equation}
\label{Tbil}
T\chi_i^+\chi_jT^{-1}
=
\begin{cases}
\chi^+_{[i+2]}\chi_{[j+2]} &
\text{if $i,j\le2 N-2$ or $i,j\ge 2N-1$},
\\[2mm]
\omega^{-1}\chi_{i+2}^+\chi_{[j+2]}\chi_{2N+1} &
\text{if $i\le 2N-2$ and $j\ge 2N-1$},
\end{cases}
\end{equation}
where the sum of the indices in the square bracket must be carried out modulo $2N$. More
precisely:  $\chi_{[j+2]}:=\chi_{1+(j+1)\text{mod}\, 2N}$. Hence, we identify: $\chi_{[2N+1]}:=\chi_1$, and
   $\chi_{[2N+2]}:=\chi_2$.
We see that the translation on a single spin node produces a shift by two steps
on the particle lattice. This property once again reflects the fact that a single spin is encoded by two
parafermions.

Note that the transformation of the time-reversal particle operators $\bar\chi_i$ is described
by the complex conjugate versions of Eqs.~\eqref{Tchi}, \eqref{Tchi-N}, \eqref{Tbil}. The only change is
the replacement of the  phase factors there by the inverted values: $\omega,\epsilon\to\omega^{-1},\epsilon^{-1}$.

The translation rules \eqref{Tchi}, \eqref{Tchi-N} for the Majorana fermions take
the following form:
\begin{equation}
\label{Tgamma}
\begin{aligned}
T\gamma_iT^{-1}=-\imath\gamma_1\gamma_2\gamma_{i+2},
\qquad
T\gamma_{2N-1}T^{-1}=-\imath\gamma_2 \gamma_{2N+1},
\qquad
T\gamma_{2N}T^{-1}=\imath\gamma_1 \gamma_{2N+1}
\end{aligned}
\end{equation}
where the range of the index $i$ is defined above \eqref{Tchi}.

\medskip

Finally, we remark that the lattice transformations of particles considered in the current article
are inherited from those for the underlying spins \eqref{par}, \eqref{T}.
This is the reason of a more complex behavior of particle operators under the  chain reflection \eqref{Pchi} and
translation \eqref{Tchi}, \eqref{Tchi-N} compared to the case when the parafermions or Majorana fermions
\cite{seiberg} are the basic particles.

\medskip

\paragraph*{Hamiltonian  and its symmetries.}
Now, we are ready to construct the parafermionic  representation of the cluster system with closed boundaries.
As was discussed earlier,  the chain's size must be  even: $N=2L$. This choice is made
in order to guarantee the  invariance of the initial spin Hamiltonian \eqref{H}--\eqref{Heven}
under the translation on two lattice sites:
\begin{equation}
\label{T2H}
T^2HT^{-2}=H.
\end{equation}
One can construct the translationally invariant
model starting from  the parafermion chain with open boundary conditions.
Indeed, from the action of the cyclic shift on the bilinears \eqref{Tbil},  it is easy to see that under the $T^2$
translation
most local terms  of the corresponding Hamiltonian \eqref{Hpar2} map to the next term with the index incremented by four:
$\chi_i\to \chi_{i+4}$. The exceptions are the four terms at the top right edge
and their Hermitian conjugates. These, according to Eq.~\eqref{Tbil},  are transformed into the additional three-body
interaction terms
$$
T^2\chi_{2N-4}^+\chi_{2N-1}T^{-2}
=\omega^{-1}\chi_{2N}^+\chi_3\chi_{2N+1},
\qquad
T^2\bar\chi_{2N-6}^+\bar\chi_{2N-3}T^{-2}
=\omega\bar\chi_{2N-2}^+\bar\chi_1\chi_{2N+1},
$$
 which supplement the open system to the closed translationally invariant model.
Thus, the
 parafermionic  cluster Hamiltonian with periodical boundary conditions is given as follows:
\begin{equation}
\label{Hpar-per-2}
H=-\frac1{2}\sum_{r=1}^{L-1}\left(\epsilon^{-1} \chi_{4r}^+\chi_{4r+3} + \epsilon\bar\chi_{4r-2}^+\bar\chi_{4r+1}\right)
-\frac1{2}\left(\epsilon\chi^+_{2N}\chi_3  + \epsilon^{-1}\bar\chi^+_{2N-2}\bar\chi_1\right) \chi_{2N+1}
   +\text{H. c.}
\end{equation}
The above Hamiltonian now contains all primary particles, and none is preserved among them in contrast to
the open chain case \eqref{comHb-even}, \eqref{comHb'}. This is in good agreement with the uniqueness of the SPT ground state
in the closed systems.

Remember now that the composite parafermion operator $\chi_{2N+1}$ is the diagonal element  of the $\mathbb{Z}_p\times \mathbb{Z}_p$ symmetry
of the Hamiltonian \eqref{chi-last}. Therefore, the entire state space is split into separate $q$-sectors,
each specified by the eigenvalue $\chi_{2N+1}=\omega^q$ where $q=0,1,\dots, p-1$. Obviously, due to the parafermion
exchange relations, any other particle operator, say, $\chi_1$ together with its integral powers implements a one-to-one
map between different $q$-sectors.
Thus, the sectors are equivalent and have dimension $p^{N-1}$. When restricted to the $q$-sector,
the Hamiltonian \eqref{Hpar-per-2} simplifies taking the bilinear form:
\begin{equation}
\label{Hpar-per-1}
H_q=-\frac1{2}\sum_{r=1}^{L-1}\left(\epsilon^{-1} \chi_{4r}^+\chi_{4r+3} + \epsilon\bar\chi_{4r-2}^+\bar\chi_{4r+1}\right)
-\frac{\omega^q}{2}\left(\epsilon\chi^+_{2N}\chi_3  + \epsilon^{-1}\bar\chi^+_{2N-2}\bar\chi_1\right)
+\text{H. c.}
\end{equation}

Clearly,  translational invariance  enhances the symmetry of the Hamiltonian \eqref{Hpar-per-2}.
Since in the current section we deal with the  even $N$ case,
the system \eqref{Hpar-per-2} remains invariant under
the lattice reflection $\mathcal{P}$ \eqref{PH1}.
Under translation by a single site, the above Hamiltonian
maps to its time reversal:
\begin{equation}
\label{Hpar2T-per}
\begin{aligned}
THT^{-1}&=\mathcal{T}H\mathcal{T}^{-1}=\bar H
\\
&= -\frac1{2}\sum_{r=1}^{L-1}\left(\epsilon\bar\chi_{4r}^+\bar\chi_{4r+3} + \epsilon^{-1}\chi_{4r-2}^+\chi_{4r+1}\right)
-\frac1{2}\left(\epsilon^{-1}\bar\chi^+_{2N}\bar\chi_3  + \epsilon\chi^+_{2N-2}\chi_1\right)\chi_{2N+1}
+\text{H. c.}
\end{aligned}
\end{equation}
 This property can be established using Eq.~\eqref{Tbil}, or is more easily observed when looking at  the spin chain
  \eqref{H}--\eqref{Heven}.
 Moreover, the joint $T\mathcal{T}$ transformation similar to the space parity $\mathcal{P}$,
 maps the even and odd parts of the Hamiltonian to each other.  Hence, it
 is a symmetry of the closed model:
\begin{equation}
\label{TT}
[T\mathcal{T}, H]=[\mathcal{P}, H]=0,
\qquad
T\mathcal{T}H_\text{even}(T\mathcal{T})^{-1}=\mathcal{P}H_\text{even}\mathcal{P}=H_\text{odd}.
\end{equation}
Of course, both transformations  mix the stabilizers inside in different ways. Notice that
the $T\mathcal{T}$ (antiunitary translation) symmetry is essential in establishing the low-energy behavior of spin
chains with staggered chiral three-spin interaction \cite{oshikawa23}.

The two operators $T\mathcal{T}$ and $\mathcal{P}$, which are subjected to the algebraic relation
$$
\mathcal{P}(T\mathcal{T})\mathcal{P}=(T\mathcal{T})^{-1},
$$ generate a symmetry group
with $2N$
elements which may be regarded as a antiunitary analog of the dihedral group
$\mathbb{D}_{N}$. Obviously, the even powers of the   generator $T\mathcal{T}$ are unitary
while its odd powers are antiunitary. In particular, $(T\mathcal{T})^N=(T\mathcal{T})^{2L}=1$, and
 $(T\mathcal{T})^2=T^2$. Therefore, the $T^2$ (the translation on two spin sites) symmetry \eqref{T2H} may be considered
 as a part of more general $T\mathcal{T}$ symmetry of the Hamiltonian  \eqref{Hpar-per-2}.

Due to the nondegeneracy of the closed model, the ground state \eqref{gs} must be invariant
under both transformation \eqref{TT}. Moreover, it forms the trial representation of
the symmetry group considered in the current article.  Actually, it was already observed that it has
an even parity quantum number \eqref{PH2} under the reflection. From the other side,
both the translation and time reversal \eqref{Tgs} acting on the vacuum state produce the same
result so  their combined action preserves it with the unit quantum number:
\begin{equation}
\label{TTgs}
T\mathcal{T}|0\rangle = |0\rangle:
\qquad
T|0\rangle =  p^{-\frac{N}2} \sum_{n_1,\dots, n_N}  \omega^{\sum_{l=1}^{N} (-1)^{l}n_l n_{l+1}} |n_N n_1\dots n_{N-1}\rangle
=\bar{|0\rangle}.
\end{equation}
The $\mathbb{Z}_p\times \mathbb{Z}_p$ symmetry, which protects the topological order
of the cluster model, also acts trivially on the translationally invariant ground state \eqref{gs}:
\begin{equation}
\label{Xgs-per}
X_\text{even}|0\rangle = X_\text{odd}|0\rangle = \chi_{2N+1}|0\rangle
={|0\rangle}.
\end{equation}
The above condition does not hold for the open SPT state  Rq.~\eqref{Xgs}.
The relations \eqref{PH2}, \eqref{TTgs}, and \eqref{Xgs-per} mean that the
vacuum state of the closed system form a trivial, or singlet representation
with respect to the whole considered symmetry group.



In case of Majorana fermions, the time reversal  \eqref{time-maj}
 becomes a  symmetry of the Hamiltonian
regardless of the boundary conditions: $\bar H=H$.
Therefore, the periodic system remains invariant under the translation on a single spin step
\eqref{Tgamma}:
\begin{equation}
\label{HMaj-per}
H=THT^{-1}=\imath\sum_{l=1}^{N-2}\gamma_{2l}\gamma_{2l+3}
-\imath\left(\gamma_{2N}\gamma_3  + \gamma_{2N-2}\gamma_1\right) \gamma_{2N+1}.
\end{equation}
 This reduces to two decoupled periodic Kitaev chains.
 There are two sectors with $q=0,1$ corresponding to  the eigenvalues $(-1)^q=\pm 1$
 of the composite Majorana operator $\gamma_{2N+1}$. The restriction of
 the above Hamiltonian on them  gives two decompled Majorana chains with antiperiodical
 and periodical boundary conditions:
\begin{equation}
\label{HMaj-per-2}
H_{q=0}=\imath\sum_{l=1}^{N-2}\gamma_{2l}\gamma_{2l+3}
-\imath\left(\gamma_{2N}\gamma_3  + \gamma_{2N-2}\gamma_1\right),
\qquad
H_{q=1}=\imath\sum_{l=1}^{N}\gamma_{2l}\gamma_{[2l+3]}.
\end{equation}
Here the square brackets in the last Majorana operator takes into account
the periodicity condition similar to the one in Eq.~\eqref{Tbil}.
It is worth mentioning that the second system with the odd fermion parity is the only case when the
spin Hamiltonian retains  the (true) translational invariance when expressed in terms of the particle
operators. The statement is easily verified  using Eq.~\eqref{Hpar-per-1} since it requires the
condition $\omega^{q-1}=1$ which is fulfilled only for the fermions in the sector with the odd particle number
\cite{LSM,seiberg}.

 Similar to the open chain case \eqref{Hpar1}, the cluster Hamiltonian also may be expressed
 exclusively in terms of the original particle operators. To accomplish this, it remains
 to write in $\chi_i$ operators the last  boundary term in Eq.~\eqref{Hpar-per-2}
 which can be done by returning to the  spin model and making use of Eqs.~\eqref{chi1-4}, \eqref{ZN}
 and the exchange relations of particle operators:
\begin{equation}
  \label{per2}
 \epsilon^{-1}\bar\chi^+_{2N-2}\bar\chi_1 \chi_{2N+1} =  Z_{N-1} X_NZ^+_1 
 =\epsilon^{-5}\chi_{2N-2}\chi^{+\,2}_{2N-1}\chi_{2N}^2\chi_1^+\chi^+_{2N+1}.
\end{equation}
As a result, an alternative parafermion representation of the closed model \eqref{Hpar-per-2} is given by the
following formula:
\begin{multline}
\label{Hpar1-per}
H=-\frac{1}{2\epsilon}
\sum_{r=1}^{L-1} \left(
\chi_{4r}^+\chi_{4r+3}
+
\omega^2\chi_{4r-2}\chi_{4r-1}^2\chi_{4r}^{+\,2}\chi_{4r+1}^+
\right)
\\
-\frac{\epsilon}{2} \left( \chi^+_{2N}\chi_3 \chi_{2N+1}+ \omega^3\chi_{2N-2}\chi^{+\,2}_{2N-1}\chi_{2N}^2\chi_1^+\chi_{2N+1}^+\right)
+\text{H. c.}
\end{multline}

\medskip
\paragraph*{String order parameters.}
Remember that the Hamiltonian \eqref{Hpar-per-2} is built from the local stabilizers, which must
satisfy the condition
\begin{equation}
\label{stab}
\epsilon^{-1}\chi_{4r}^+\chi_{4r+3}|0\rangle
=\epsilon\bar\chi_{4r-2}^+\bar\chi_{4r+1}|0\rangle
=\epsilon^{-1}\chi_3^+\chi_{2N}|0\rangle
= \epsilon\bar\chi_1^+\bar   \chi_{2N-2}|0\rangle
=|0\rangle,
\qquad r=1,2,\dots, \frac N2-1,
\end{equation}
where the $\chi_{2N+1}$ is excluded for the last two equations \eqref{Xgs-per}.
Thus, any product of above operators also stabilizes the SPT state.
In particular, the nonlocal string order operator is composed of the product of $m$ sequential even or odd
local stabilizers \eqref{string}.
Their parafermionic representations are also nonlocal:
\begin{align}
\label{str-e}
Z_{2r-1}X_{2r}X_{2r+2}\dots X_{2r+2m-2}Z^+_{2r+2m-1} &=
\epsilon^{-m}
\prod_{r'=r}^{r+m-1}\bar\chi_{4r'-2}^+\bar\chi_{4r'+1},
\\
\label{str-o}
Z^+_{2r}X_{2r+1}X_{2r+3}\dots X_{2r+2m-1}Z_{2r+2m}
&=
\epsilon^{m}
\prod_{r'=r}^{r+m-1}\chi_{4r'}^+\chi_{4r'+3}.
\end{align}
Note that the above operators   can be easily modified to the case when the string
passes through the edges of the periodic system. Evidently, all these quantities
mutually commute, and in case of  $m=1$, they are reduced
to the local Hamiltonians.

Similar to the Hamiltonian
and  its $\mathbb{Z}_p\times\mathbb{Z}_p$
symmetry generators, the string parameters are distributed between the two subsystems.
Namely, the operator \eqref{str-o} is associated with the odd chain which is described in terms of the
$\chi$ parafermions while the operator \eqref{str-e} is associated with the even chain given by the
$\bar\chi$ particles.

\section{Summary}

The current article is devoted to the parafermionic representation of the
cluster chain, which possesses  the $\mathbb{Z}^\text{odd}_p\times \mathbb{Z}^\text{even}_p$
symmetry-protected
topological (SPT) order.
It corresponds to the $p$-state extension of the spin cluster with $p=2$ and
reduces to the noninteracting Potts model in  the trivial phase.

The  Fradkin-Kadanoff transformation maps the cluster
model into one-dimensional Hamiltonian with parafermions ($\chi_i$) and highly nonhomogeneous
(odd and even) local terms.
We have derived the action of time reversal (${\mathcal T}$),  lattice parity (${\mathcal P}$)
and translation ($T$)
on particle operators.
Furthermore, we have established that the use of time-reversal
particle operators ($\bar\chi_i$) restores a similarity between odd and even terms.
As a result, we obtained   two independent chains bilinear in particle operators.
The first chain is composed
of standard parafermions ($\chi_{4r}$, $\chi_{4r+3}$),  while the second one
consists of time-reversal parafermions ($\bar\chi_{4r-2}$, $\bar\chi_{4r+1}$).
These subsystems respectively inherit the $\mathbb{Z}^\text{odd}_p$
and $\mathbb{Z}^\text{even}_p$ components of the symmetry, which describe the total parity of particles
contributing to each chain.
Under the parity-conservation condition, the topological phases
become intrinsic,
and the long-range  SPT order parameters turn into the topological order parameters
of the $\chi$- and $\bar\chi$-chains.
This finding establishes the correspondence between the SPT phase of the
cluster model and the topological phases of two parafermionic systems. Furthermore, we extend this correspondence
to include the boundary modes.

Specifically, it has been demonstrated that under open boundary conditions,
the independent zero-energy edge excitations are equally redistributed along the $\chi$- and $\bar{\chi}$-chains.
We have highlighted four invariant parafermionic modes located at the edges of each chain,
which are associated with the boundary invariants of the spin system. These boundary modes generalize the Majorana zero-modes of Kitaev’s superconducting chain.
At the same time, the  structure of these  zero-modes differs
for even-length and odd-length  chains. This difference arises due to the distinct
symmetries present in both systems:
 the first model  is $\mathcal{P}$-invariant
 while the second one is  $\mathcal{PT}$-invariant.

We also consider the translationally invariant model with an even number of spins and derive its space-time symmetries.
An additional $T\mathcal{T}$ symmetry emerges, which maps the $\chi$- and $\bar{\chi}$-chains to each other.
Together with $\mathcal{P}$ symmetry, it forms an antiunitary analog of the dihedral group.
Consequently, the corresponding Hamiltonian becomes twisted by the total parafermion parity operator,
which is analogous to Dirac's chiral $\gamma$ matrix.
The eigenvalues of the parafermion parity are $p$-th roots of unity.
In a sector with a specific eigenvalue, the Hamiltonian becomes bilinear in parafermions with twisted boundary conditions.

In the Ising case ($p=2$),  the parafermions together with their time reversals
turn into the Majorana fermions. Then, the spin cluster splits up into two  separate
Kitaev chains in the topological phase as has been established recently  \cite{verresen}.

\acknowledgments
The authors are grateful to Wen-Tao Xu and Thomas Quella for useful remarks and comments.
T. H. thanks  also Samuel Elman for interesting discussion.
The research was supported by the Armenian Science Committee grants Nos. 21AG-1C047 and 24FP-1F039.

\end{document}